\documentclass[12pt,preprint]{aastex}

\shorttitle{Discovery of an eccentric H$\alpha$ emitting disk in M31}
\shortauthors{Menezes et al.}

\begin{document}

\title{Discovery of an H$\alpha$ emitting disk around the supermassive black hole of M31}

\author{R. B. Menezes, J. E. Steiner, and T. V. Ricci}

\affil{Instituto de Astronomia Geof\'isica e Ci\^encias Atmosf\'ericas, Universidade de S\~ao Paulo, Rua do Mat\~ao 1226, Cidade Universit\'aria, S\~ao Paulo, SP CEP 05508-090, Brazil;}
\email{robertobm@astro.iag.usp.br}

\begin{abstract}

Due to its proximity, the mass of the supermassive black hole in the nucleus of Andromeda galaxy (M31), the most massive black hole in the Local Group of galaxies, has been measured by several methods involving the kinematics of a stellar disk that surrounds it. We report here the discovery of an eccentric H$\alpha$ emitting disk around the black hole at the center of M31 and show how modeling this disk can provide an independent determination of the mass of the black hole. Our model implies a mass of $5.0_{-1.0}^{+0.8} \times 10^7 M_{\sun}$ for the central black hole, consistent with the average of determinations by methods involving stellar dynamics, and compatible (at 1$\sigma$ level) with measurements obtained from the most detailed models of the stellar disk around the central black hole. This value is also consistent with the $M-\sigma$ relation. In order to make a comparison, we applied our simulation on the stellar kinematics in the nucleus of M31 and concluded that the parameters obtained for the stellar disk are not formally compatible with the parameters obtained for the H$\alpha$ emitting disk. This result suggests that the stellar and the H$\alpha$ emitting disks are intrinsically different from each other. A plausible explanation is that the H$\alpha$ emission is associated with a gaseous disk. This hypothesis is supported by the detection of traces of weaker nebular lines in the nuclear region of M31. However, we cannot exclude the possibility that the H$\alpha$ emission is, at least partially, generated by stars.

\end{abstract}


\section{Introduction}

All massive galaxies appear to host a supermassive black hole (with $10^6-10^{10} M_{\sun}$) at their center \citep{gul09,mcc11}. Measuring the mass of central black holes in galaxies is of great importance, as the discovery of a relationship between mass and the velocity dispersion of the stars in the central bulge, the $M-\sigma$ relation \citep{fer00,geb00}, reveals the possible co-evolution of black holes and their host galaxies \citep{gra04}.

M31, the Andromeda galaxy, is an Sb galaxy at a distance of 778 kpc and its nucleus can be observed with excellent spatial resolutions. \citet{lig74}, using data obtained with the Stratoscope II, revealed an asymmetry in the nuclear region of M31, as the bright nucleus did not coincide with either the center of the bulge or the maximum of the stellar velocity dispersion. However, \citet{lau93}, using observations from the \textit{Hubble Space Telescope} (\textit{HST}), showed that the galaxy possesses a double nucleus, the two components being called P1 (the brightest one) and P2 (located, approximately, at the center of the bulge). These two components are separated by about $0\arcsec\!\!.49$. 

A model to explain the morphology of the nucleus of M31 was proposed by \citet{tre95} and states that P1 and P2 are parts of an eccentric stellar disk around the black hole, with P1 coinciding with the apocenter and the black hole being located at P2. Several refinements to this model have been put forth \citep{sal04,pei03}; \citet{ben05}, using \textit{HST} data, revealed that the black hole is actually located in a structure embedded in P2 called P3, which probably corresponds to a cluster of A-type stars. \citet{lau12}, using also \textit{HST} data, confirmed that P3 corresponds to a cluster of blue stars around the central black hole.

The mass of the central black hole of M31 has already been measured by, at least, six different techniques: (1) standard dynamical modeling ignoring asymmetries \citep{dre88,kor88}; (2) the center of mass argument, which depends on the asymmetry of P1+P2 \citep{kor99}; (3) dynamical modeling of the stellar nuclear disk taking into account the asymmetry of P1+P2 \citep{pei03}; (4) complete dynamical modeling taking into account the asymmetries and the self-gravity of the nuclear stellar disk of P1+P2 \citep{sal04}; (5) dynamical modeling of P3, which is independent of P1+P2 \citep{ben05}; (6) \textit{N}-body simulations \citep{bac01}. All of these methods involved stellar dynamics and resulted in values in the range $0.3 - 23 \times 10^7 M_{\sun}$ for the mass of the central black hole in M31.

In this Letter, we analyze a data cube of the nuclear region of M31, obtained with the Integral Field Unity (IFU) of the Gemini Multi-Object Spectrograph (GMOS) of the Gemini North telescope, and report the discovery of an eccentric H$\alpha$ emitting disk around the central black hole.

\section{Observations, reduction and data treatment}

The observations of M31 were made on 2009 September 21. We used the IFU of the GMOS of the Gemini North telescope, in the one-slit mode, in order to obtain data cubes, with two spatial dimensions and one spectral dimension. The science field of view (FOV) has $5\arcsec \times 3\arcsec\!\!.5$, while the sky FOV (observed simultaneously at a distance of $1\arcmin$ from the science FOV) has $5\arcsec \times 1\arcsec\!\!.75$. Three 10 minute exposure of the nuclear region of M31 were made, with the grating B600-G5307, in a central wavelength of $6000\AA$. The final spectra had a coverage of $4550 - 7415\AA$ and a resolution of $R\sim 2900$. The estimated seeing for the night of observation was $0\arcsec\!\!.55$.

Standard calibration images were obtained during the observations. The data reduction was made in IRAF environment. At the end of the process, three data cubes were obtained, with spaxels of $0\arcsec\!\!.05 \times 0\arcsec\!\!.05$. No sky subtraction was applied because the sky FOV (still inside the disk of M31) was contaminated with stellar emission from the galaxy.

After the data reduction, we performed a procedure of data treatment. First, a correction of the differential atmospheric refraction was applied to all data cubes, using an algorithm developed by our group. In order to combine the three corrected data cubes into one, a median of these data cubes was calculated. After that, a Butterworth spatial filtering \citep{gon02}, with order $n = 2$, was applied to all the images of the resulting data cube, in order to remove spatial high-frequency noise. Finally, a Richardson-Lucy deconvolution \citep{ric72, luc74} was applied to all the images of the data cube, using a synthetic Gaussian point-spread function (PSF). The PSF of the final data cube has FWHM $\sim 0\arcsec\!\!.45$. 

\begin{figure}
\plotone{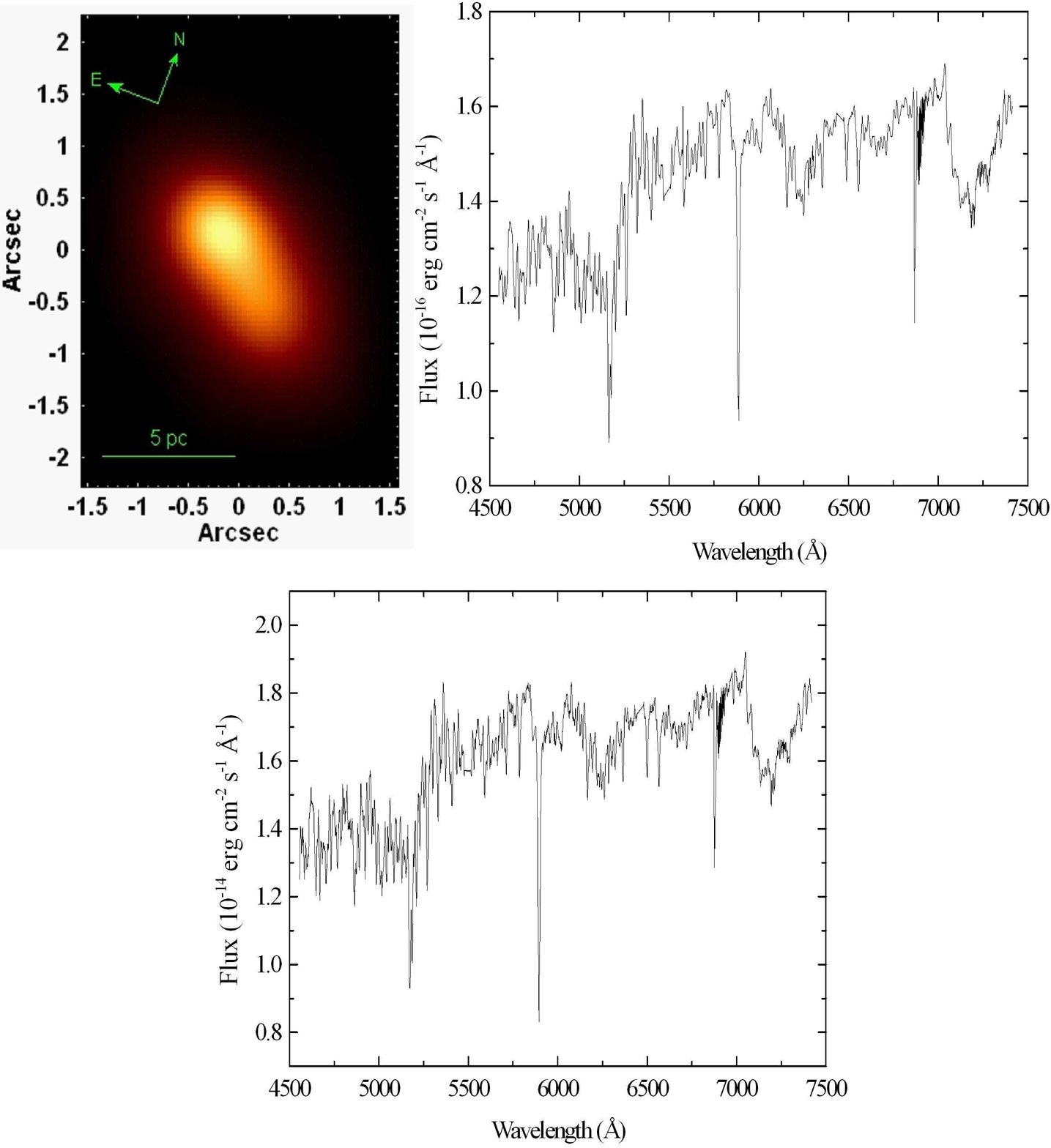}
\caption{Left above: image of the final data cube of M31 collapsed along the spectral axis. Right above: average spectrum of the final data cube of M31. Bottom: spectrum extracted from a circular area, with a radius of $0\arcsec\!\!.15$, centered on P1.\label{fig1}}
\end{figure}

Figure~\ref{fig1} shows an image of the final data cube of M31 (obtained after the data treatment) collapsed along the spectral axis and an average spectrum of this data cube. The brightest component of the nucleus, P1, can be easily detected; however, the fainter components, P2 and P3, cannot be seen, due to the spatial resolution and to the lack of spectral sensitivity in the blue (below $5000\AA$). A spectrum of P1, extracted from a circular area with a radius of $0\arcsec\!\!.15$, is also shown in Figure~\ref{fig1}. The average signal-to-noise ratio (S/N), between $5610\AA$ and $5670\AA$, of the spectra of the data cube analyzed here is close to 50.

\section{Data analysis and results}

After the data treatment, a spectral synthesis was applied to the spectrum of each spaxel of the resulting data cube of M31. This procedure was performed with the Starlight software \citep{cid05}, which fits the stellar spectrum of a given object with a combination of template stellar spectra from a pre-established base. In this work, we used the base of stellar spectra MILES (Medium resolution INT Library of Empirical Spectra; S\'anchez-Bl\'azquez et al. 2006). The spectral synthesis resulted in a synthetic stellar spectrum for each spaxel. These synthetic spectra were then subtracted from the observed ones, leaving a data cube with emission lines only. The non subtraction of the sky field during the data reduction had no observable effect in the results obtained with the spectral synthesis. 

In this residual data cube, we have detected a previously unreported weak extended H$\alpha$ emission. The luminosity of this H$\alpha$ emission line, within a radius of $0\arcsec\!\!.7$ from the central black hole, is $L_{H\alpha}(emission) \approx (8.7 \pm 1.0) \times 10^2 L_{\sun}$; the equivalent width of this line is $W_{H\alpha}(emission) \approx -0.18 \pm 0.03\AA$, while that of the absorption component is $W_{H\alpha}(stellar) \approx 2.03 \pm 0.10\AA$, indicating that the extended H$\alpha$ emission is much weaker than the stellar component and, therefore, hard to detect. Figure~\ref{fig2} shows the spectral fit of the H$\alpha$ absorption line in one spectrum of the data cube of M31 and also the fit residuals. It is possible to detect traces of an H$\alpha$ emission line immersed in the absorption component, however, only the subtraction of the spectral fit allowed a clear visualization of this very weak emission line. Figure~\ref{fig2} also reveals the presence of H$\beta$ and [O III]$\lambda5007$ emission lines. The spectral resolution of the base of stellar spectra MILES (FWHM = $2.3\AA$) is very similar to our spectral resolution (FWHM = $2.2\AA$) and, because of that, the spectral fit is quite precise and allows an accurate subtraction of the stellar emission (Figure~\ref{fig2}). The spectra of the data cube of M31 were resampled to $\Delta\lambda = 1\AA$ (the same spectral sampling of MILES spectra), before the spectral synthesis was applied.

In Figure~\ref{fig2}, we can see a map of the integrated flux of the H$\alpha$ emission line. Only the central region of the FOV of the data cube was taken into account here, because the H$\alpha$ emission was too faint farther away. We can see two bright areas in the vicinity of P1 and P2, respectively. This reveals a certain similarity between the extended H$\alpha$ emission and the double nucleus of M31 and suggests the existence of a possible H$\alpha$ emitting disk. An elongated emitting region from SE toward NW can also be seen; however, we do not know the origin of this peculiarity in the morphology of the extended H$\alpha$ emission. The area with masked values in the map corresponds to an emitting region, observed in previous studies \citep{del00}, that is not associated to the extended H$\alpha$ emission detected here. This emitting region can be identified by observing its intense, compact, and narrow [O III]$\lambda\lambda4959,5007$ emission.

In order to obtain a velocity map for H$\alpha$, we fitted a Gaussian function to the H$\alpha$ emission line in each spectrum of the data cube. Figure~\ref{fig2} shows an example of the Gaussian fit applied to the H$\alpha$ emission line detected in one spectrum of the data cube, after the subtraction of the spectral fit. We can see that, despite the irregularities of the emission line, the Gaussian fit provides (with a considerable precision) the wavelength corresponding to the peak of the emission line and, therefore, the radial velocity. On the other hand, the irregularities observed in the H$\alpha$ emission lines, after the subtraction of the spectral fit, made impossible to determine reliable values for the velocity dispersion. Therefore, such values were not taken into account in this work. 

The velocity map obtained for the H$\alpha$ emission line is shown in Figure~\ref{fig3} and it clearly indicates the presence of an H$\alpha$ emitting disk around the central black hole. The position angle of the line of nodes of this emitting disk is P.A.$_e$ = $59\degr\!\!.3 \pm 2\degr\!\!.7$, which is very close to the value measured for the stellar disk in previous studies (P.A.$_l$ = $56\degr\!\!.4 \pm 0\degr\!\!.2$) \citep{bac01}. The contours in Figure~\ref{fig3} show that the ascending node of the H$\alpha$ emitting disk is very close to P1, while P2 and P3 are located in an area with low velocities. This behavior is very similar to what is observed in the stellar disk. We extracted a velocity curve along the line of nodes (Figure~\ref{fig3}) and the modulus of the maximum and minimum velocities are considerably different, indicating that any kinematical model should be eccentric. 

We tried to reproduce the velocity curve and the velocity map of H$\alpha$ using a model of a simple eccentric disk around the supermassive black hole. Only a region within a radius of $0\arcsec\!\!.7$ from the black hole was modeled because, as mentioned before, the disk was too faint farther away. We admitted that the stellar mass inside the radius of the modeled area was small compared to the mass of the black hole, so no stellar mass was taken into account in the model (see more details below). The emitting disk was simulated by superposing 33 concentric Keplerian orbits with the following free parameters: the argument of the pericenter $\omega$, the inclination of the disk $i$, the eccentricity of the disk $e$ and the mass of the central black hole $M_{\bullet}$. We measured the value of the longitude of the ascending node in the velocity map ($\Omega = 128\degr\!\!.3$), so this parameter was fixed in our model. In each simulation, after all the orbits were superposed, the model was convolved with the estimated PSF, in order to simulate the effect of the Earth's atmosphere. The free parameters were varied and the simulations were made repeatedly, in order to minimize the value of the $\chi^2$ between the observed velocity map and the simulated one:
\\
\begin{equation}
\chi^2=\frac{1}{I}\sum_{i=1}^{N_x}\sum_{j=1}^{N_y}\frac{w_{ij}\cdot I_{ij}\cdot \left(v_{ij}(observed)-v_{ij}(simulated)\right)^2}{\sigma_{ij}^2},
\end{equation}
\\
where $N_x$ is the number of spaxels along the horizontal axis, $N_y$ is the number of spaxels along the vertical axis, $\sigma_{ij}$ is the uncertainty of the velocity of the spaxel $(i,j)$, $v_{ij}(observed)$ is the velocity of the spaxel $(i,j)$ of the observed velocity map, $v_{ij}(simulated)$ is the velocity of the spaxel $(i,j)$ of the simulated velocity map, $w_{ij}$ is weight equal to 1 for areas near the line of nodes and equal to 0 for farther areas, $I_{ij}$ is the value of the spaxel $(i,j)$ of the map of integrated fluxes of H$\alpha$ (Figure~\ref{fig2}), and $I$ is the sum of the values of all spaxels of the map of integrated fluxes of H$\alpha$ (Figure~\ref{fig2}).

In the previous formula, the weight $w_{ij}$ is represented by a step function, which is equal to 1 for spaxels closer than $0\arcsec\!\!.25$ to the line of nodes and equal to 0 for farther spaxels. We used the values of the integrated flux of the H$\alpha$ emission line ($I_{ij}$) in the calculation of the $\chi^2$ in order to give more weight to the spaxels with higher S/N.

The best simulated velocity map and the corresponding best simulated velocity curve are also shown in Figure~\ref{fig3}. The parameters of the best model obtained, with $\chi^2 = 1.67$, are shown in Table~\ref{tbl1}. The uncertainties (1$\sigma$) of the parameters were estimated by plotting histograms (probability distributions) of each one of the parameters of the simulation, considering, only cases in which $\chi^2 - \chi^2_{min} < 1$. After that, we fitted a Gaussian function on each histogram and obtained the square deviation, which was taken as the uncertainty of each parameter \citep{der11}. We can see that, despite some irregularities of the observed velocity map, our model of a simple eccentric disk reproduced the observed kinematical behavior considerably well. In Figure~\ref{fig3}, however, we can see a clear discrepancy between the observed velocity map and the simulated one in an area near the ascending node. This is the same emitting region that was masked in the map of the integrated flux of the H$\alpha$ emission line (Figure~\ref{fig2}), which is not associated to the eccentric disk detected here. A by-product of this modeling was an independent determination of the position of the black hole. We found that it is at a distance of $0\arcsec\!\!.050 \pm 0\arcsec\!\!.025$ from P2, which is compatible (at 1$\sigma$ level) with the position of P3 (at a distance of $0\arcsec\!\!.076$ from P2) from previous determinations \citep{bac01}.

As mentioned before, in this model, we have not taken into account the effect of the mass of the stars. In order to evaluate this assumption, we used an \textit{HST} image of the nucleus of M31 in \textit{V} band, obtained with WFPC2, to estimate the stellar mass within the radius of the simulated area. First, we decomposed this image into an asymmetric component (containing the stellar disk around the black hole) and a symmetric one (containing the central part of the stellar bulge). After that, we de-projected the symmetric component using a multi-Gaussian expansion approach \citep{cap02}. Finally, the stellar masses of the two components were calculated using a mass-to-light ratio of $M/L_{V} = 5.7$ \citep{ben05}. The results are $M_{sym} = 8.7 \times 10^5 M_{\sun}$ for the bulge component and $M_{asym} = 8.7 \times 10^6 M_{\sun}$ for the disk. Together, these two components represent $19 \%$ of the $M_{\bullet}$ we obtained. That means that this value of $M_{\bullet}$ resulting from our model can be, at most, $19 \%$ too high, due to the neglect of the stellar mass. Introducing this in the previous uncertainty for the black hole mass, we obtain a final value of $M_{bh\_H\alpha} = 5.0_{-1.0}^{+0.8} \times 10^7 M_{\sun}$. Since the spectral synthesis performed on the data cube provided details about the stellar populations detected, we could, in principle, have used a mass-to-light ratio derived from these results to estimate the stellar mass in the nuclear region of M31. However, we decided to use a mass-to-light ratio derived from \textit{HST} observations \citep{ben05} because, due to our spatial resolution and lack of spectral sensitivity in the blue (below $5000\AA$), we did not observe certain young stellar populations in the nucleus of M31 that were clearly detected in \textit{HST} observations.

We performed an entirely analogous modeling on a stellar velocity map obtained from the data cube of M31. Only a region within a radius of $0\arcsec\!\!.8$ from the black hole was modeled because the effect of the self-gravity of the stellar populations could be considerable at farther areas. The parameters of the best model obtained are shown in Table~\ref{tbl1} and the final mass obtained for the central black hole (including the uncertainty introduced by the neglect of the stellar mass in the modeling) is $M_{bh\_stellar} = 4.5_{-1.1}^{+0.9} \times 10^7 M_{\sun}$. We will show the details about this topic in R. B. Menezes et al. (2013, in preparation), since the stellar kinematics analysis is not the purpose of this Letter.

\section{Discussion and conclusions}

The mass of the central black hole in M31 we obtained by modeling the kinematics of the H$\alpha$ emission as a simple eccentric disk ($M_{bh\_H\alpha} = 5.0_{-1.0}^{+0.8} \times 10^7 M_{\sun}$) is within the range of the values found in previous determinations using stellar kinematics ($10^{6.5} - 10^8 M_{\sun}$ in \citet{kor88}; $3-7 \times 10^7 M_{\sun}$ in \citet{dre88}; $3.3_{-1.8}^{+1.2} \times 10^7 M_{\sun}$ in \citet{kor99}; $1.0 \times 10^8 M_{\sun}$ in \citet{pei03}; $1.4_{-0.3}^{+0.9} \times 10^8 M_{\sun}$ in \citet{ben05}; $7.0_{-3.5}^{+1.5} \times 10^7 M_{\sun}$ in \citet{bac01})  and it is compatible (at 1$\sigma$ level) with the measurement ($5.62 \pm 0.66 \times 10^7 M_{\sun}$) obtained with the most detailed models of the stellar disk elaborated so far \citep{sal04}. This result is also compatible (at 1$\sigma$ level) with the value of the mass of the central black hole obtained by modeling the stellar kinematics as a simple eccentric disk ($M_{bh\_stellar} = 4.5_{-1.1}^{+0.9} \times 10^7 M_{\sun}$). Most of the recent models used to reproduce the stellar disk around the black hole in M31 did not use a single inclination and eccentricity for the entire disk. However, the values we obtained for $i$ and $e$ in our best models are in the range of the ones used in several studies to model the stellar disk ($0 - 0.7$ for the eccentricity and $41\degr - 77\degr$ for the inclination) \citep{sal04,tre95,pei03}. 

\begin{figure}
\epsscale{0.79}
\plotone{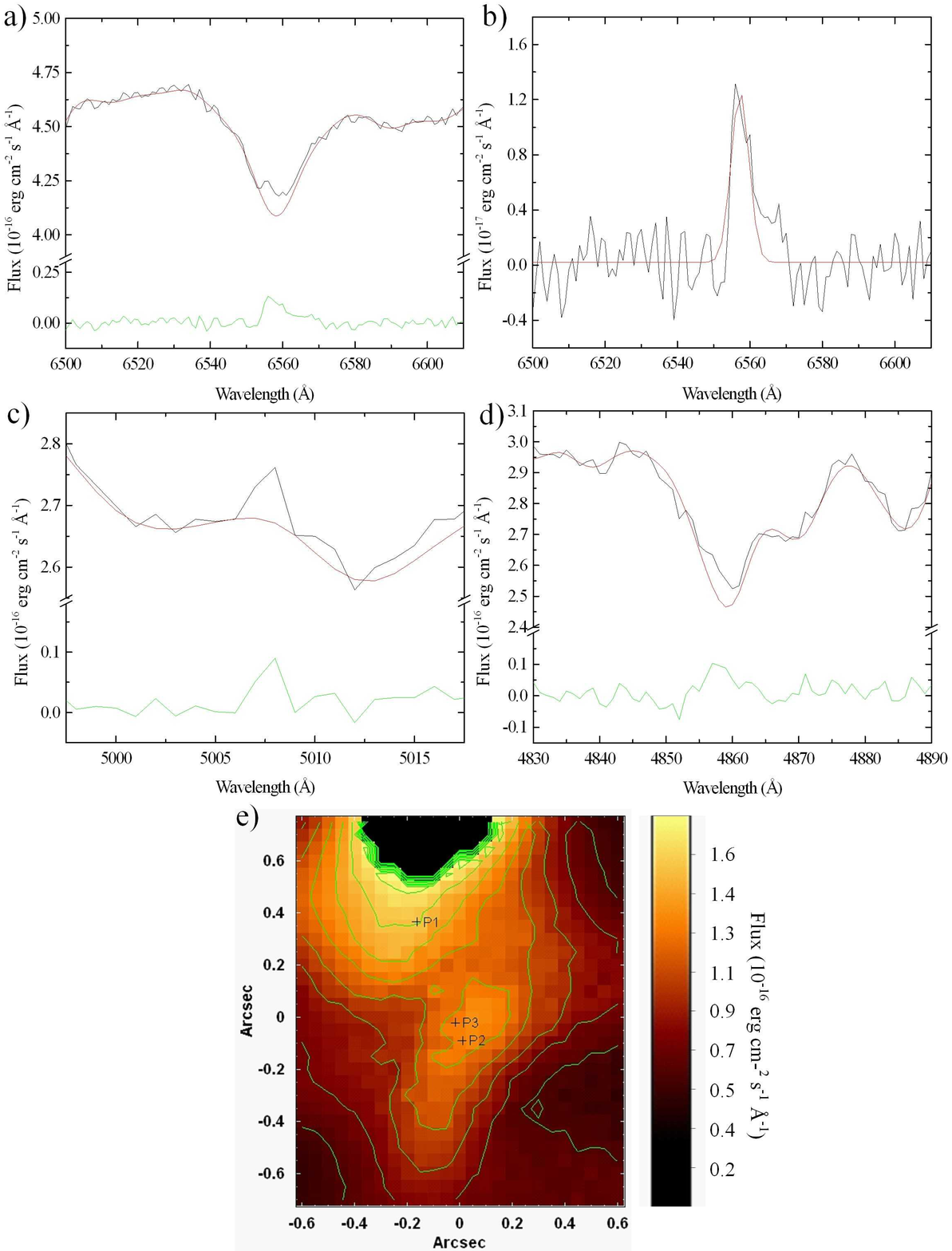}
\caption{a) H$\alpha$ absorption line of one spectrum of the data cube of M31 ($+0\arcsec\!\!.1$;$-0\arcsec\!\!.3$), with the spectral fit (in red) and the fit residuals (in green). b) Same fit residuals shown in a), with a Gaussian fit (in red). c) [O III] $\lambda5007$ emission line of the same spectrum shown in a), with the spectral fit (in red) and the fit residuals (in green). d) H$\beta$ absorption line of the same spectrum shown in a), with the spectral fit (in red) and the fit residuals (in green). e) Map of the integrated flux of the H$\alpha$ emission line. The positions of P1, P2, and P3 and the contours are also shown.\label{fig2}}
\end{figure}

\begin{figure}
\epsscale{1.0}
\plotone{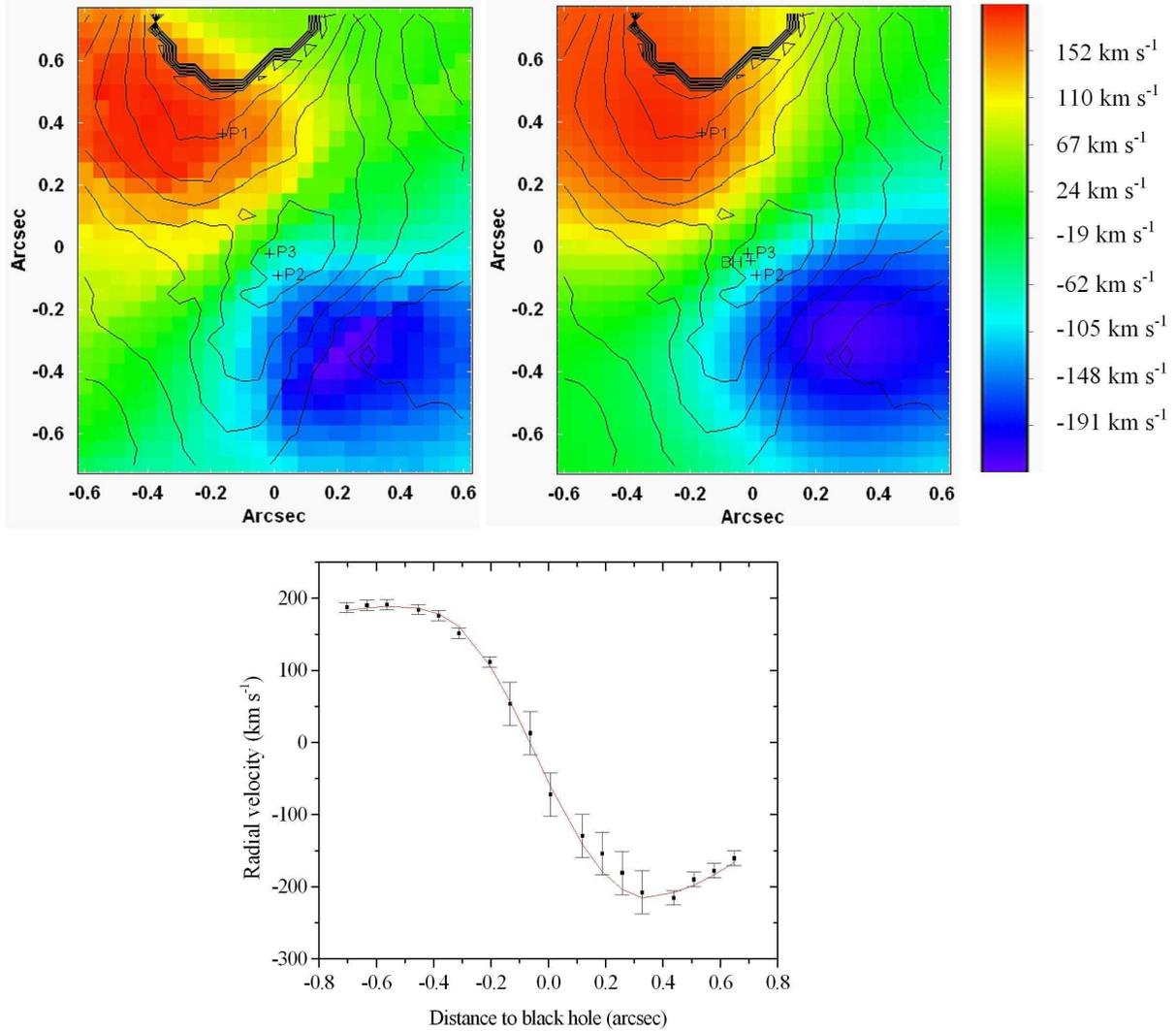}
\caption{Left above: observed velocity map of the H$\alpha$ line. Right above: best simulated velocity map of the H$\alpha$ line, with the position of BH (the black hole in our model) indicated. The contours of the integrated flux of the H$\alpha$ line (Figure~\ref{fig2}) and the positions of P1, P2, and P3 are shown in the observed and in the best simulated velocity maps of H$\alpha$. Bottom: velocity curve of the H$\alpha$ emission line, measured along the line of nodes. The best simulated velocity curve (in red) is superposed on the measurements (points with error bars).\label{fig3}}
\end{figure}

The values of $e$ and $i$ obtained for the stellar and the H$\alpha$ emitting disks with our models are compatible at 1$\sigma$ level. This reveals a certain similarity between these two models. However, the values of $\omega$ obtained are not compatible at all. This suggests that the stellar and the H$\alpha$ emitting disks are intrinsically different from each other. According to what was mentioned before, we have identified a weaker H$\beta$ emission line and also traces of even weaker nebular lines, like [N II] $\lambda6583$, [S II] $\lambda\lambda6716,6731$, and [O III]$\lambda\lambda4959,5007$ (Figure~\ref{fig2}), along the FOV of the residual data cube of M31. This, together with the intrinsic difference between the stellar and the H$\alpha$ emitting disks, suggests that the H$\alpha$ emission is associated with a gaseous disk. However, we cannot exclude the possibility that this emission is, at least in part, originated by stars.

The discovery of an H$\alpha$ emitting disk in the nucleus of M31, reported here, allows an independent measurement of the mass of the central black hole in M31, and, therefore, has a considerable importance for the studies of this object.

\acknowledgments

This work is based on observations obtained at the Gemini Observatory. We thank FAPESP for support under grants 2008/11087-1 (R.B.M.) and 2008/06988-0 (T.V.R) and also an anonymous referee for valuable comments about this Letter.

{\it Facilities:} \facility{Gemini:Gillett(GMOS)}.

\clearpage

\clearpage

\clearpage

\clearpage

\begin{table}
\begin{center}
\caption{Parameters of the Best Model Obtained for the H$\alpha$ Emitting Disk and for the Stellar Disk.\label{tbl1}}
\begin{tabular}{ccc}
\tableline\tableline
Parameter & H$\alpha$ Emitting Disk & Stellar Disk \\
\tableline
$M_{\bullet}$ & $(5.0\pm0.8) \times 10^7 M_{\sun}$ & $(4.5\pm0.9) \times 10^7 M_{\sun}$\\
$e$ & $0.35\pm0.07$ & $0.30\pm0.12$ \\
$\omega$ & $115\degr\pm12\degr$ & $250\degr\pm17\degr$ \\
$i$ & $45\degr\pm4\degr$ & $55\degr\pm10\degr$ \\
\tableline
\end{tabular}
\end{center}
\end{table}

\end{document}